\documentclass[conference, table]{IEEEtran}
\usepackage[colorinlistoftodos, textwidth=2cm,textsize=tiny]{todonotes}
\usepackage{xspace}
\usepackage{paralist}

\usepackage{algorithm2e}
\usepackage[noend]{algpseudocode}
\usepackage{amsmath}
\usepackage{amssymb}
\usepackage[normalem]{ulem}

\usepackage[show]{chato-notes}

\usepackage{booktabs}
\usepackage{multirow}

\usepackage{balance} 

\usepackage[compress, sort]{cite} 

\usepackage{graphicx}
\usepackage{float}
\usepackage[caption = false]{subfig}

\usepackage{url}

\usepackage[export]{adjust box}
\usepackage{pgf}

\usepackage{hyperref}
\usepackage{setspace}

\newcommand{\spara}[1]{\smallskip\noindent{\bf #1}}

\newcounter{hypothesis}

\makeatletter
\def\@IEEEpubidpullup{8\baselineskip}
\makeatother
\begin{document}

\title{
    Modeling Human Annotation Errors to Design \\ 
    Bias-Aware Systems for Social Stream Processing   
}

\author{
\IEEEauthorblockN{Rahul Pandey}
\IEEEauthorblockA{\textit{George Mason University} \\
Fairfax, USA \\
rpandey4@gmu.edu}
\and
\IEEEauthorblockN{Carlos Castillo}
\IEEEauthorblockA{\textit{Universitat Pompeu Fabra} \\
Barcelona, Spain \\
carlos.castillo@upf.edu}
\and
\IEEEauthorblockN{Hemant Purohit}
\IEEEauthorblockA{\textit{George Mason University} \\
Fairfax, USA \\
hpurohit@gmu.edu}
} 

\maketitle





\begin{abstract}

High-quality human annotations are necessary to create effective machine learning systems for social media. 
%
Low-quality human annotations 
indirectly contribute to the creation of inaccurate or biased learning systems. 
We show that human annotation quality is dependent on the ordering of instances shown to annotators (referred as `annotation schedule'), and can be improved by local changes in the instance ordering provided to the annotators, yielding a more accurate annotation of the data stream for efficient real-time social media analytics. 

We propose an error-mitigating active learning algorithm that 
is robust with respect to some cases of human errors when deciding an annotation schedule. 
We validate the human error model 
and evaluate the proposed algorithm against strong baselines 
by experimenting on classification tasks of relevant social media posts during crises. 
%
According to these experiments, considering the order in which data instances are presented to human annotators leads to both an increase in accuracy for machine learning and awareness toward some potential biases in human learning that may affect the automated classifier.  
%
%
%

\end{abstract}




\section{Introduction and Background}\label{sec:intro}

\begin{figure}[t]
\centering
\includegraphics[width=0.9\linewidth,trim=2 2 2 2,clip]{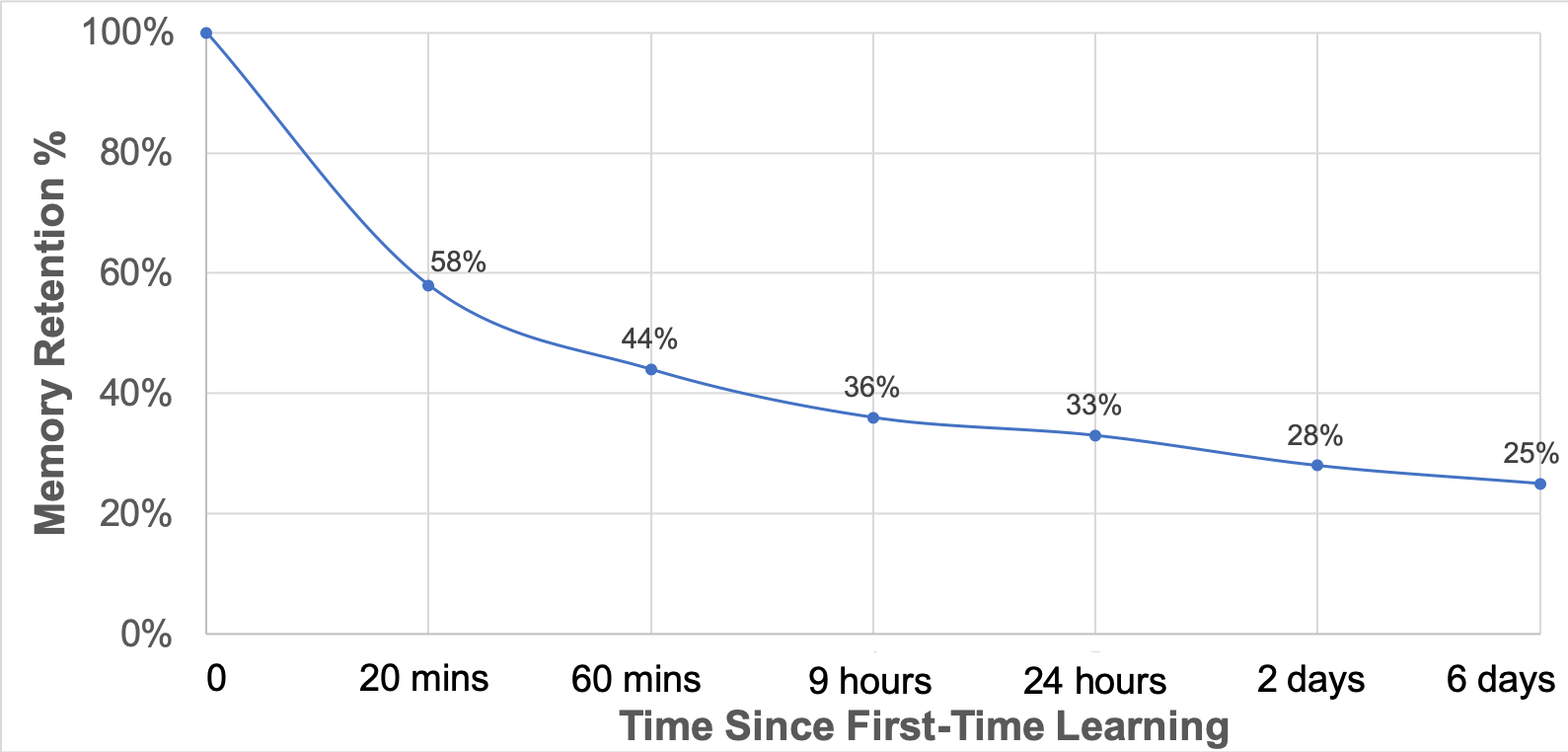}
\caption{The Ebbinghaus Curve for forgetting behavior of humans, as described in a 2013 paper~\cite{ebbinghaus2013memory}.  
We investigate the effect of forgetting on the human annotation quality for HSPS and corresponding mitigation approaches. 
\label{fig:forgetting-curve}}
\vspace{-0.3in}
\end{figure}

Filtering of high-volume, high-velocity social media streams is a typical task in many domains such as journalism, public health, and crisis management. 
In this scenario, an avalanche of data must be filtered and classified to prevent information overload. 
This is challenging as these data streams are often noisy, sparse, and redundant. 
They require a substantial cognitive effort to be processed by humans, who cannot scale. 
These data streams are also problematic for pure machine annotation systems, as depending on the task they may have limited accuracy due to the data challenges such as concept drifts and domain adaptation. 
%
Hence, to achieve high accuracy 
the active learning framework is suitable 
%
%
to design hybrid stream processing systems (HSPS) through a composition of human annotation and automatic classification~\cite{imran2013engineering,lofi2014design}. 
The scope of this paper is to study a hybrid online classification setting that categorizes relevant instances from a social stream. 
Specifically, we analyze the effect of human forgetting (see Figure~\ref{fig:forgetting-curve})  errors in such annotation tasks.

\spara{Human challenges in hybrid stream processing.} 
%
Systems that rely on some form of crowdsourcing are known to be affected by the cognitive biases and cognitive load of human annotators~\cite{burghardt2018quantifying}.   
The first key factor is annotator burnout, as high workload causes a deterioration of the quality of annotations~\cite{marshall2013experiences}. 
To prevent burnout, one can cap the maximum number of annotation tasks per unit of time that the annotator must perform, which can reduce workload~\cite{purohit2018ranking}.  
The second key factor is human error in the execution of annotation task. A popular error taxonomy proposed by Reason~\cite{reason1990human} suggests that recurrent error forms may have their origins in psychological processes. 
We particularly study two types of errors, \textit{mistakes} and \textit{slips}~\cite{reason1990human,zhang2004cognitive}:  
\emph{Mistakes} are errors due to incorrect or incomplete knowledge~\cite{reason1990human}, and in the annotation context correspond to annotators who have not yet grasped the concept to be annotated, or who are annotating new instances for which they have not acquired a correct representation yet. 
\emph{Slips} are errors in the presence of correct and complete knowledge~\cite{reason1990human}, i.e., the annotator knows the correct category for an instance but selects an incorrect one. If this error becomes persistent after a large number of examples, 
 then it may indicate burnout. 


\spara{Online active learning.}   
Existing types of online active learning methods, to the best of our knowledge, address only possible machine biases and errors.  
Gama et al.~\cite{gama2014survey} and Almeida et al.~\cite{almeida2018adapting} provide extensive surveys of the different active learning methods. The primary categories for the active learning methods include the one group focused on the better sampling of the instance space for querying (e.g., addressing concept drift~\cite{zliobaite2014active}), and another group focused on better learning of a discriminatory model.        
However, the improvement of both of the above active learning methods for stream processing systems requires to account the mitigation for the potential oracle/human errors during querying as well, to be efficient and unbiased in the modeling.

\spara{Our contribution.}
We formalize a theoretical-motivated human error typology of \textit{mistakes} and \textit{slips} that covers some common types of human errors possible in a stream annotation task. 
We demonstrate the validity of the error typology using forgetting error model, which is tested via lab and crowdsourcing annotation experiments for the information filtering task in crisis datasets. 
We then present a novel method for human error-mitigation in active learning for stream processing tasks against strong baselines (Section~\ref{sec:solution}) and novel insights on automatic approaches to prevent human errors. 
The application of the proposed human error framework can be used to design human-AI collaboration strategies and improve machine learning performance in hybrid stream processing systems. 



\spara{Data and task.} We define the specific annotation task for human error testing and mitigation as to classify an instance from a given sequence/stream of Twitter messages (tweets) into $k$ classes. We use labeled datasets from prior work in crisis informatics that contains labeled tweets related to natural disasters~\cite{alam2018crisismmd}. We recrawled the tweet instances from Twitter  API to acquire the metadata such as timestamp and also, to discard any tweets deleted since the data was originally collected. The two disaster events include 
Hurricanes \textit{Harvey} 
and \textit{Irma}. The labels 
include four tweet categories: 
\textit{infrastructure and utility damage} ($c1$) - information about any physical damage on infrastructure or utilities; \textit{rescue, volunteering, and donation effort} ($c2$) - information about offering help through volunteering efforts by a community of users; \textit{affected individuals} ($c3$) - information about the condition of the individuals during this disaster event, and \textit{not relevant or cannot judge} ($c4$). 
We considered human labels with confidence score (computed by the crowdsourcing platform for agreement between multiple annotators~\cite{alam2018crisismmd}) greater than 65\%. 


\section{Lab-scale Annotation Error Testing\label{sec:lab-experiments}} 
We focus on quantifying the forgetting behavior~\cite{ebbinghaus2013memory}, which underlies the above-mentioned error types and impacts the performance of both human annotation and machine learning. 


\subsection{Forgetting Curve} 
Psychologists have been studying human forgetting behavior in the context of learning and acquiring new knowledge for centuries. The Ebbinghaus Curve -- shown in Figure~\ref{fig:forgetting-curve} -- is a famous experimental result for forgetting in humans~\cite{ebbinghaus2013memory}. 
Inspired by this curve, we hypothesize that the forgetting behavior of humans can be approximately modeled by a simple function; 
we use a sigmoid function given the similar asymptotic nature of it and forgetting curves. 
We quantitatively model the error rate as a function of the time during which a human annotator has not seen any instance of a given class to annotate in the stream. We define \emph{forgetting\_score} for a class $c$ over time $t$ lapsed after its last annotated instance as: 
\begin{equation}
    \operatorname{forgetting\_score}(c) = \gamma \cdot \frac{1}{1+e^{-\alpha t + \beta}}
    \label{eqn:forgetting_score}
\end{equation}

\subsection{Experimental Validation}
For validating the pattern of human forgetting behavior through the above function, we conducted a small-scale lab study in a controlled environment with 3 human annotators. We set up an environment where each user has to classify a message from an input stream into one of four classes. Messages included a random amount of irrelevant messages (noise) between the ground truth messages, to better observe human forgetting of the class and human errors. 
Our data stream contains 800 samples; 3 annotators label a given instance in the stream one by one, with no ability to go back. Once we collect the responses from the three annotators, we calculate the \emph{forgetting\_score} for each instance of the 800 samples based on ground truth data for the class labels of each instance and when the class instance last appeared in the instance sequence. Thus, we observe the annotator responses whether or not the annotators actually forget that class. 


Our experiment validated the modeling of forgetting behavior using the sigmoid function that we hypothesized (plots omitted for brevity). This motivates us to use sigmoid function based forgetting model to induce an error in our algorithmic simulation experiments to mimic the real environment. 





 

\section{Simulation-based Error Testing \& Mitigation}\label{sec:solution} 
We simulate the annotation task in an active learning scenario for online streams~\cite{zliobaite2014active}. We design a novel method for generating a dynamic annotation schedule for an annotator (simulated ``oracle'') such that the schedule attempts to minimize human errors and maximize the overall performance of the active learning algorithm. 
\subsection{Mitigation Algorithms}

%
Our method first samples a batch of $n$ instances from a time interval $t_i$ by using a conventional uncertainty sampling algorithm for ``oracle'' annotation (in our proposed  approach, we also apply constraints to select only $m$ ($m<n$) instances that minimize the human forgetting error). We then update the learning model using the annotated instances for predictions in the next time interval $t_{i+1}$. 
For simulation, we use the ground truth labels 
as the oracle annotations. 
We propose three types of algorithms (first two are baselines) based on diverse sampling strategies to select instances at the end of $t_i$: 

\spara{(Baseline) Algorithm 1: Random Sampling}. We randomly sample a batch of
$n$ instances (equal to the number of samples in the uncertainty region, as described next) 
at interval $t_i$. We hypothesize that random sampling can address the issue of data distribution changes for the concept drift by selecting an instance from any region in the concept space, although it may be inefficient to improve the learning performance over time. 

\spara{(Baseline) Algorithm 2: Uncertainty Sampling}. We predict the classes of incoming batch of instances with the current active learning model (created at time interval $t_{i-1}$). 
After prediction, we select the classified instances in which the prediction confidence is in the range of $[30\%, 70\%]$ as the uncertainty region ($n$ instances.) We provide these uncertainty region instances to the oracle and get their annotations. 
Our hypothesis is that the model will become more robust if it starts learning from the cases on the decision boundary region.  

\spara{(Proposed) Algorithm 3: Error-mitigating Sampling}. This algorithm 
relies on uncertainty sampling to first select candidates from uncertain region. It then discards the instances whose predicted class (from the model at $t_{i-1}$) could either induce performance errors to the new model at $t_i$ 
or tend to be not forgotten by the oracle (i.e., the bias in the human learning behavior toward that class). 

Specifically, for each instance $X_y$ in the uncertain region, we predict its class based on our current model and check if it is not the class to discard. Otherwise, we select that instance for annotation by the oracle and update our model. 
%
We store the instance, its arrival time, and annotation to vectors 
that contain information for all instances of the context window (the most recent $l$ intervals: ($t_{i-l}$,$t_i$)), in order to compute the class to discard next. 
Likewise, we maintain and append  
an error matrix $E^{a\times b}$ where $a$ is the total instances from the current context window and $b$ is the total number of class pairs ($|C|\times (|C|-1)$ ) minus the same class pairs. 
The error matrix column corresponds to the class pairs, i.e. $(c_2, c_1), (c_3, c_2)$, etc. and a row corresponds to the information for an instance $X_y$.  For every new instance $X_y$ annotated with $c_k$ that is added to the error matrix, a matrix cell $E(X_y,(c_j, c_k))$ represents the error in predicting the instances annotated with $c_j$ in the past context window, using the model updated by including current instance $X_y$ of class $c_k$. 
All remaining cells of the row $X_y$ 
are copied from the previous row ($X_{y-1}$) of the error matrix in the previous state. 
We use the error matrix $E$ to decide the classes to discard only after the first three intervals.  

%
%
%

We calculate the bias score for each class $c_k\in C$ using error matrix rows for all the instances $X_y$ annotated with $c_k$ as: 
\begin{center}
\vspace{-0.22in}
    \begin{equation}
        \operatorname{BiasScore} (c_k) = \sum_{\forall X_y}  \;\; \sum_{c_j \in |C|-c_k} E(X_y, (c_j,c_k))
    \end{equation}
\vspace{-0.1in}
\end{center}

We also add forgetting factor to the bias scores for each class that signifies the time lapsed since the last time an instance of the class was observed in the stream. 

The forgetting factor is quantitatively defined as:  
\begin{center}
\vspace{-0.22in}
    \begin{equation}
        \operatorname{ForgetScore(c_k)} = e^{-\Delta T_k}
    \end{equation}
    where $\Delta T_k = $ time difference from the last two appearance of class $c_k$ instances in the context window 
\vspace{-0.1in}
\end{center}

The final score for each of class $c_k$ 
is defined as: 
\begin{center}
\vspace{-0.22in}
    \begin{equation}
        \operatorname{Score} (c_k) = \operatorname{ForgetScore} (c_k) \times \operatorname{BiasScore} (c_k)
    \end{equation}
\vspace{-0.2in}
\end{center}
Once we calculate the final scores for each class $c_k$, we determine the class $c_k$ with the highest score as the error-inducing class to discard at current time. 

\subsection{Simulation Experiments\label{sec:simulations}} 

We describe the data preparation for the simulated stream processing task and the active learning environment. 

\spara{Data Preparation.} 
We split the data into train, test, and warm-up sets. The 20\% of the whole data is used as a test set. From the remaining 80\%, we randomly picked $z$ instances ($z = 20$) of each class to create a warm-up set; the rest constitutes the train set. Since we have a class imbalance in our data, we use an equal number of instances across classes for creating our warm-up phase model for robustness. 
The training data is sorted based on the arrival time of an instance (tweet) in the stream. After sorting, we divided the data into equal bins of size $N$. At each interval, $N$ instances would arrive for annotation and get filtered for inclusion in the train set based on our mitigation algorithms ($N>n>m$). The $N$ is fixed and is computed in the beginning as follows:  
\vspace{-0.2in}
\begin{center}
\begin{equation} 
    N = \frac{\operatorname{train\_set\_size}}{(\operatorname{train\_time}_{max} - \operatorname{train\_time}_{min})_{days}}
    \label{formulae:N}
\end{equation}
where $train\_time_{min}$ and $train\_time_{max}$ are the least and highest timestamp of instances in the train set (unit: days). 
\end{center}
The reason to keep $N$ fixed is due to the fact that our labeled dataset is not continuous in real-time setting but is distributed along a long time span, given it was annotated through crowdsourcing method in prior work. Hence, we cannot fix $N$ based on time units (such as seconds or minutes).  

\spara{Active Learning Environment.} 
 We implement the active learning process following previous work~\cite{zliobaite2014active}. First, we train the base model with our warm-up set and then, keep updating. 
\begin{figure*}[ht!]
 \centering
\includegraphics[width = 2.33in]{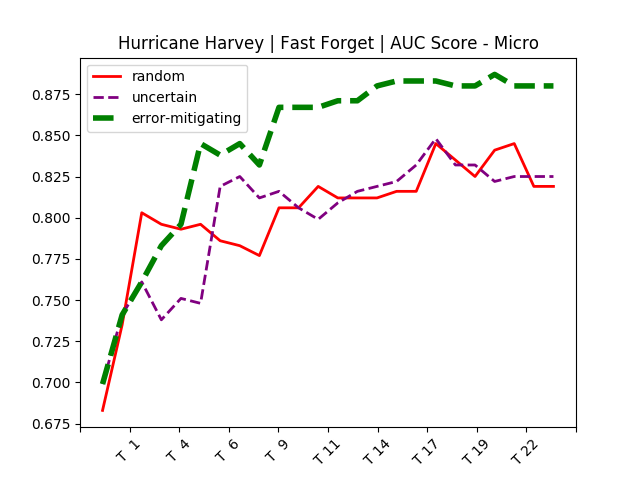} 
\includegraphics[width = 2.33in]{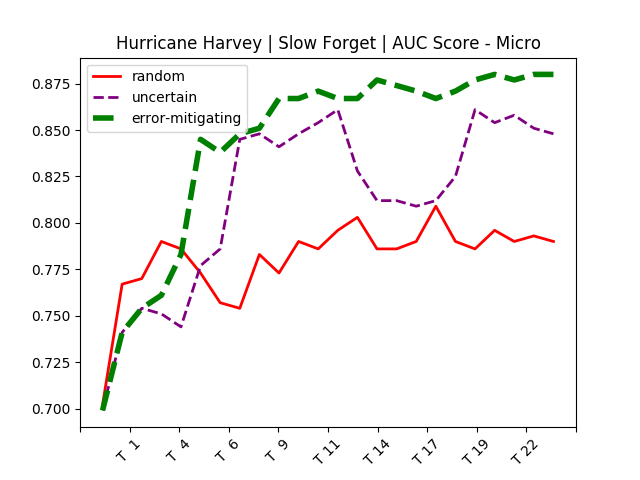} 
\includegraphics[width = 2.33in]{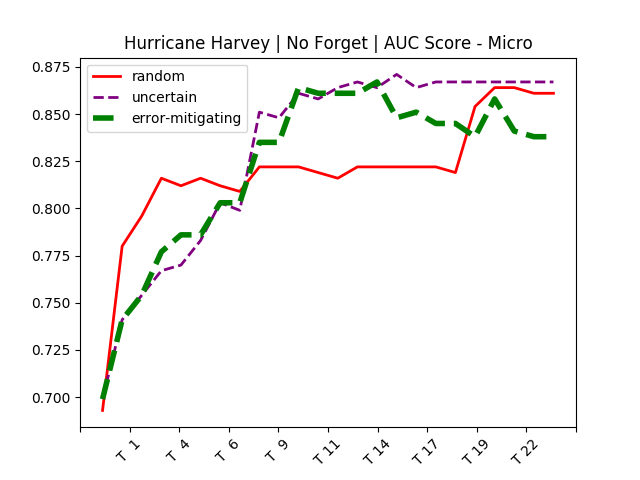} \\ 
\includegraphics[width = 2.33in]{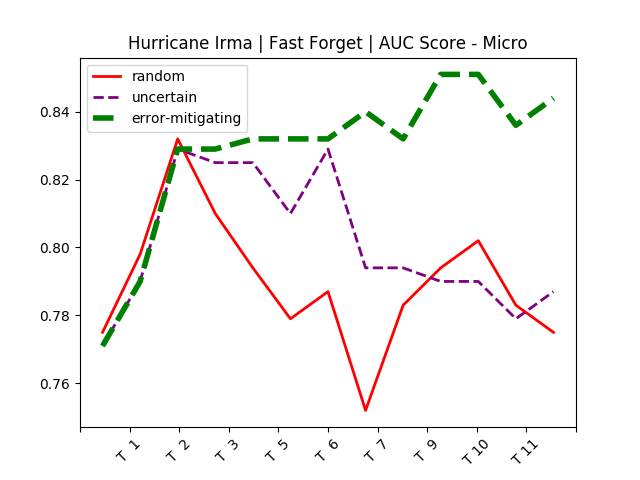} 
\includegraphics[width = 2.33in]{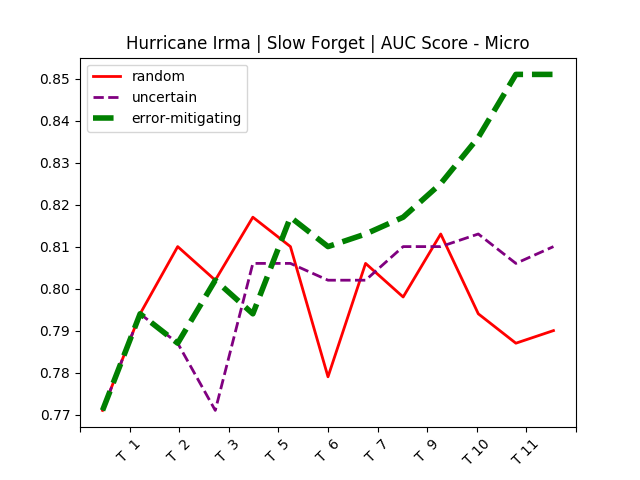} 
\includegraphics[width = 2.33in]{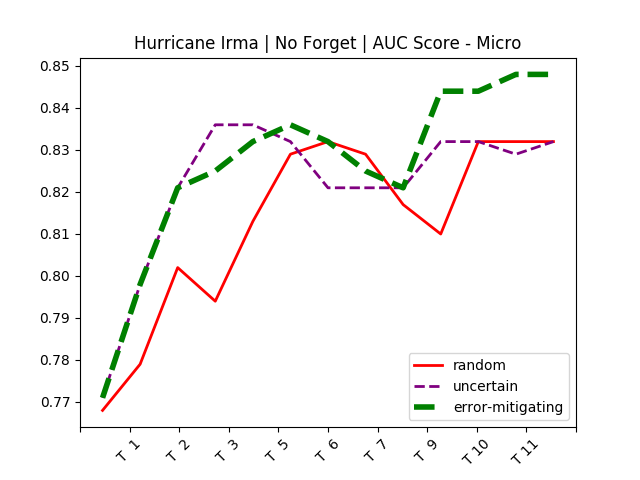} \\ 
 \caption{AUC score of mitigation algorithms for hurricane datasets, showing superior performance of error-mitigating sampling in the case of forgetting errors.  
} 
\label{fig:auc-micro}
\vskip -0.1in
\end{figure*} 
We used pretrained GloVe-Twitter embeddings 
 with 200 dimensions for 
 word-level features and then, averaged them 
 to represent tweet-level features. We train a linear SVM model 
 and measure the performance 
 on the fixed test set. 

For every interval $t_i$, we receive $N$ instances for seeking the annotator feedback in order to acquire more labeled data for retraining the current model. Depending upon the mitigation algorithms, i.e., random, uncertain, or error-mitigating sampling, we filter the $n$ or $m$ instances to get annotations from the oracle. To mimic human behavior, the label given by the oracle is not always correct. Based on the lab-scale experimental results of Section~\ref{sec:lab-experiments}, the forgetting behavior of humans follows the sigmoid function from our annotation task. 
Thus, we utilize the value of a sigmoid function with different parameters to find the probability that the oracle generates a correct or erroneous label due to forgetting the class as given in the equation \ref{eqn:forgetting_score}. 
We use 3 different parameter settings to add errors through the forgetting behavior of the oracle: 
    (1) \textit{Slow Forgetting}: computes a sigmoid function with parameters estimated from errors in a human experiment: $\alpha = 0.0434$, $\beta = 0.9025$, and $\gamma = 0.75$.
    (2) \textit{Fast Forgetting}: uses a sigmoid function that converges to $1$ faster than the slow forgetting and induces errors more frequently: 
    $\alpha = 0.03$, $\beta = 1.00$, and $\gamma = 1.00$.
    (3) \textit{No Forgetting}: assumes that our oracle always gives the correct labels and does not forget any class. Hence, we use the true labels for each annotation.

\section{Discussion and Conclusion}\label{sec:discussion}
\vspace{-0.04in}
We note that when we try to mimic the real world scenario by inducing different types of forgetting errors (slow vs. fast forgetting) in the simulated annotation process, 
our error-mitigating sampling algorithm is able to reduce the effect of human forgetting and improve AUC scores over time across all the event datasets, despite varying amount of instances per interval.  
Also, for the first three intervals, the simple uncertainty-based and our  error-mitigating algorithms perform similarly. It is possible due to 
insufficient learning to enable the decision making for discarding a class that is likely going to induce errors to other classes. 
Both these algorithms show a gradual increment of performance as the new instances arrive, as compared to the random sampling algorithm with highly variant behavior of learning. These observations support that our proposed algorithm could help improve active learning systems in a real-time setting. 


In the case of error-mitigation algorithm, the chances of inducing human error are lower as it accounts for the expectation of a class forgetting likelihood, which reduces expected errors in annotating instances of such a class. 
%

%
%
%
%
The application of this method 
can help in designing human-AI collaboration systems for efficient stream processing for social media and web in general. Such systems would require not only less human annotations, but also have fewer errors and less bias from the human annotators.  

\spara{Limitations and future work.} 
This study does not cover all types of human error in stream annotation. Instead, 
it can provide a foundation to systematically study diverse types of errors and causes. 
We designed the annotation task for only text classification to test the human errors but future work can explore other types 
such as image object recognition. 

\spara{Reproducibility.} Human annotations and code implementations are available upon request for research purposes. 

\section{Acknowledgement} 
Purohit thanks U.S. NSF grant awards 1815459 \& 1657379 and Castillo thanks La Caixa project (LCF/PR/PR16/11110009) for partial support. 

\balance

\balance


\balance
\bibliographystyle{IEEEtran}
\bibliography{paper-asonam-humanerror}



\end{document}